\documentclass[aps,prb,twocolumn,groupedaddress,showpacs]{revtex4} 

\usepackage{graphicx}
\usepackage{dcolumn}
\usepackage{bm}

\begin{document} 

\title{Heat transport of La$_{2-y}$Eu$_y$CuO$_4$ and 
La$_{1.88-y}$Eu$_y$Sr$_{0.12}$CuO$_4$ single crystals} 

\author{X. F. Sun} 
\email[]{ko-xfsun@criepi.denken.or.jp} 
\author{Seiki Komiya} 
\author{Yoichi Ando} 
\affiliation{Central Research Institute of Electric Power 
Industry, Komae, Tokyo 201-8511, Japan.} 

\date{\today} 

\begin{abstract} 

To study the rare-earth doping effect on the phonon heat 
transport of La-based cuprates and shed light on the mechanism of 
phonon scattering, both $ab$-plane and $c$-axis thermal 
conductivities ($\kappa_{ab}$ and $\kappa_c$) are measured for 
La$_{2-y}$Eu$_y$CuO$_4$ ($y$ = 0, 0.02 and 0.2) and 
La$_{1.88-y}$Eu$_y$Sr$_{0.12}$CuO$_4$ ($y$ = 0 and 0.2) single 
crystals. It is found that the phonon peak (at 20 -- 25 K) in 
$\kappa_{ab}(T)$ of La$_{2-y}$Eu$_y$CuO$_4$ shows an anomalous 
Eu-doping dependence: it completely disappears for $y$ = 0.02, 
which is discussed to be due to the local lattice distortions 
around Eu dopants, and reappears for $y$ = 0.2 with much smaller 
peak magnitude compared to $y$ = 0 sample. In contrast to the 
strong suppression of the phonon peak in Eu-doped La$_2$CuO$_4$, 
Eu-doping to La$_{1.88}$Sr$_{0.12}$CuO$_4$ enhances the low-$T$ 
phonon heat transport that results in the reappearance of the 
phonon peak in this charge-carrier-doped system. The data clearly 
show that the establishment of static stripe phase rather than 
the structural change is responsible for the reappearance of 
phonon peak. 

\end{abstract} 


\maketitle 

\section{Introduction} 

In La$_{2-x}$Sr$_x$CuO$_4$ (LSCO) related compounds, the 
high-temperature structure is tetragonal (HTT, space group 
$I4/mmm$) and there are three different low-temperature 
phases:\cite{Crawford,Buechner} the LTO (low-temperature 
orthorhombic, space group $Bmab$) phase in LSCO, the LTO2 
(low-temperature orthorhombic 2, space group $Pccn$) phase in 
rare-earth (R) doped La$_{2-x-y}$R$_y$Sr$_x$CuO$_4$, and LTT 
(low-temperature tetragonal, space group $P4_2/ncm$) phase in 
La$_{2-x-y}$R$_y$Sr$_x$CuO$_4$ with larger $x$. These 
low-temperature structures are classified \cite{Crawford} by the 
tilts of CuO$_6$ octahedra around [110] and [1$\bar{1}$0] axes of 
the HTT phase and have pronounced influence on the physical 
properties. For example, it has recently been realized that the 
holes in the high-$T_c$ cuprates mesoscopically segregate into 
quasi-one-dimensional antiphase domain boundaries between 
antiferromagnetically ordered Cu spin regions, which is the 
so-called stripe phase;\cite{Tranquada1, Yamada, Mook, Wakimoto, 
Matsuda, Fujita, Hunt, Ando, Noda, Zhou, Emery, Kivelson, Zaanen, 
Ando1} the static stripe order of charges and spins is only 
established in R (Nd,Eu)-doped 
LSCO,\cite{Tranquada1,Hunt,Nachumi,Klauss,Suh} where the stripes 
are believed to be pinned by the particular tilting of CuO$_6$ 
octahedra in the LTT phase.\cite{Tranquada1} 

It is known that the antiferromagnetic (AF) insulating compound 
La$_2$CuO$_4$ (LCO), which has LTO phase, shows predominant 
phonon heat transport at low temperatures, which is manifested in 
a large phonon peak at 20 -- 25 K in the temperature dependence 
of both $ab$-plane and $c$-axis thermal conductivities 
($\kappa_{ab}$ and $\kappa_c$);\cite{Nakamura1,Sun} such phonon 
peak was found to be completely suppressed in LSCO with $x$ = 
0.10 -- 0.20.\cite{Nakamura1} Interestingly, it was found that in 
R-doped LSCO, such as La$_{1.28}$Nd$_{0.6}$Sr$_{0.12}$CuO$_4$, 
the low-$T$ thermal conductivity is much enhanced in the 
non-superconducting LTT phase, compared to that in LSCO with the 
same Sr content.\cite{Baberski} This cannot happen if the defect 
scattering and electron scattering of phonons are the only source 
of the phonon peak suppression in LSCO. Alternatively, it was 
proposed that the strong suppression of phonon heat transport in 
LSCO is due to the strong phonon scattering by the structural 
distortion associated with the dynamical stripes, while R-doping 
leads to the formation of static stripes that significantly 
reduces the phonon scattering.\cite{Baberski} However, this 
supposition would be too straightforward, because the low-$T$ 
structure phases are LTO and LTT in LSCO and R-doped LSCO, 
respectively, which possess different phonon spectra and phonon 
scattering. In fact, it has already been reported that the 
structural symmetry does affect the phonon properties; for 
example, the sound velocities in the three structures are known 
to be $v_s$(LTT) $>$ $v_s$(LTO2) $>$ $v_s$(LTO),\cite{Sera1} 
which implies that the phononic thermal conductivity is likely 
the largest in the LTT phase and the smallest in the LTO phase. 
Apparently, it is not a solid conclusion that the reappearance of 
phonon peak in R-doped LSCO is due to the formation of static 
stripes, until evidence showing a direct relationship between the 
low-$T$ phonon heat transport and the static charge stripes 
(rather than the structural transition) is obtained. Another 
question, apart from the structural phase transition, is how the 
R dopants themselves affect the phonon heat transport. A previous 
study on the Nd-doped LCO (without charge doping) has shown that 
the phonon transport is somehow suppressed by 
Nd-doping;\cite{Sera2} however, no clear R-doping dependence of 
the heat conductivity was reported, probably because only 
polycrystalline samples were used in that study. It is therefore 
desirable to clarify the role of R-doping in the phonon heat 
transport by studying both R-doped LCO and R-doped LSCO single 
crystals. 

To reduce the possible effect of magnetic moments of rare-earth 
ions on the heat transport behavior, we select Eu ion as the 
dopant, which has the smallest magnetic moment among rare-earth 
ions (atomic number 57 -- 71), except for La and Lu. In this 
paper, we report our study of the Eu-doping effect on the phonon 
heat transport in two single-crystal systems: 
La$_{2-y}$Eu$_y$CuO$_4$ ($y$ = 0, 0.02 and 0.2) and 
La$_{1.88-y}$Eu$_y$Sr$_{0.12}$CuO$_4$ ($y$ = 0 and 0.2), which 
have LTO2 and LTT phases for $y$ = 0.2, respectively, at low 
temperature. It is found that for LCO slight Eu-doping ($y$ = 
0.02) induces anomalous wipeout of the phonon peak in 
$\kappa_{ab}$, which is due to the local structural distortions 
induced by the local LTO2 regions around Eu ions in the LTO 
phase, while further increase of Eu content ($y$ = 0.2) recovers 
the phonon peak, although the peak magnitude is much smaller than 
that of undoped LCO. These results clearly show that rare-earth 
doping strongly suppress the phonon heat transport in LCO. On the 
contrary, the Eu-doping in LSCO enhances the low-$T$ phonon 
transport, which, based on the new features found in our single 
crystals, we discuss to be most likely related to the formation 
of static stripes.  

\section{Experiments} 

The single crystals of La$_{2-y}$Eu$_y$CuO$_4$ and 
La$_{1.88-y}$Eu$_y$Sr$_{0.12}$CuO$_4$ are grown by the 
traveling-solvent floating-zone (TSFZ) technique and carefully 
annealed.\cite{Komiya} After the crystallographic axes are determined 
by using the X-ray Laue analysis, the crystals are cut into 
rectangular thin platelets with the typical sizes of $2.5 \times 0.5 
\times 0.15$ mm$^3$, where the $c$ axis is perpendicular or parallel 
to the platelet with an accuracy of 1$^{\circ}$. 
La$_{2-y}$Eu$_y$CuO$_4$ samples are annealed in flowing pure He gas to 
remove the excess oxygen. On the other hand, 
La$_{1.88-y}$Eu$_y$Sr$_{0.12}$CuO$_4$ samples are annealed at 850 
$^{\circ}$C for 48 hours in air, followed by rapid quenching to room 
temperature, to remove the oxygen defects. 

The thermal conductivity $\kappa$ is measured using a 
conventional steady-state technique at 2 -- 150 K and a modified 
steady-state technique at 150 -- 300 K.\cite{Sun} The temperature 
difference $\Delta$$T$ on the sample is measured by a 
differential Chromel-Constantan thermocouple, which is glued to 
the sample using GE7031 varnish. The $\Delta$$T$ varies between 
0.5\% and 2\% of the sample temperature. To improve the accuracy 
of the measurement at low temperatures, $\kappa$ is also measured 
with ``one heater, two thermometer" method from 2 to 20 K by 
using a chip heater and two Cernox chip 
sensors.\cite{Ando2,Ando3} The errors in the thermal conductivity 
data are smaller than 10\%, which is mainly caused by the 
uncertainties in the geometrical factors. Magnetization 
measurements are carried out using a Quantum Design SQUID 
magnetometer. 

\section{Results and discussion} 

\subsection{Thermal conductivity of La$_{2-y}$Eu$_y$CuO$_4$} 

\begin{figure} 
\includegraphics[clip,width=7.8cm]{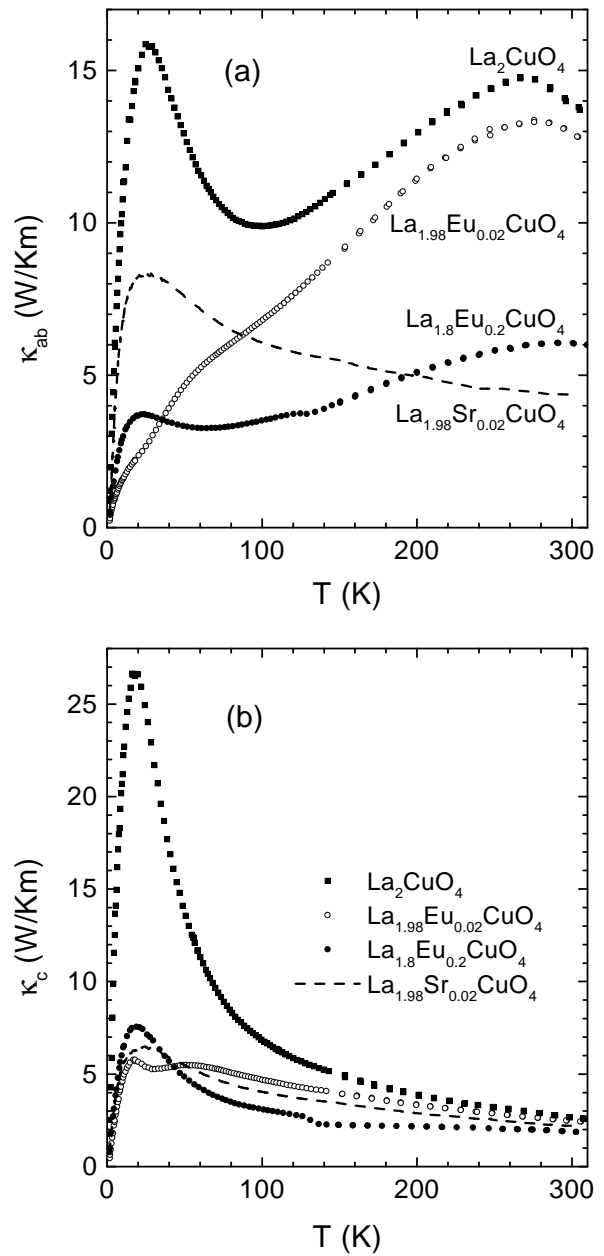} 
\caption{Thermal conductivity of La$_{2-y}$Eu$_y$CuO$_4$ ($y$ = 0, 
0.02 and 0.2) single crystals along (a) the $ab$ plane and (b) the $c$ 
axis. The data of La$_{1.98}$Sr$_{0.02}$CuO$_4$ single crystal are 
also shown for comparison.} 
\end{figure} 

Figure 1 shows the temperature dependences of the thermal 
conductivity measured along the $ab$ plane and the $c$ axis in 
pure LCO and Eu-doped LCO single crystals. The undoped 
La$_2$CuO$_4$ sample, as has already been discussed in a previous 
paper,\cite{Sun} shows a sharp phonon peak at low temperature 
[$\sim$25 K in $\kappa_{ab}(T)$ and $\sim$20 K in $\kappa_c(T)$] 
and a broad magnon peak in $\kappa_{ab}(T)$ at high temperature 
($\sim$270 K).  

Upon Eu doping, there are drastic changes in both $\kappa_{ab}$ 
and $\kappa_c$. Let us first discuss the changes of the high-$T$ 
magnon peak in $\kappa_{ab}(T)$. This peak is suppressed 
gradually upon Eu-doping, which is directly related to the 
suppression of the N\'{e}el transition temperature by Eu doping. 
The detailed magnetic behaviors of these samples are shown in the 
next subsection. 

The phonon peak at low temperatures shows an anomalous change 
with Eu doping. For La$_{1.98}$Eu$_{0.02}$CuO$_4$, with only 
1\%-Eu-substitution, the phonon peak in $\kappa_{ab}(T)$ is 
completely suppressed although the magnon peak shows only slight 
decrease. One natural phonon scattering process is caused by the 
point defects associated with dopants. However, it is found that 
the suppression of the phonon peak is much weaker in 
La$_{1.98}$Sr$_{0.02}$CuO$_4$ with the same amount of dopants, as 
shown in Fig. 1(a), which suggests that the impurity atoms 
themselves are not the main source of such strong phonon 
scattering in La$_{1.98}$Eu$_{0.02}$CuO$_4$. Another possible 
mechanism may be the magnetic scattering of phonons since 
Eu$^{3+}$ ions have magnetic moments and may disturb phonons by 
magnetoelastic coupling.\cite{Cordero,Lavrov1} If this is the 
case, some magnetic field dependence of thermal conductivity is 
expected (at least in the temperature region near the phonon peak 
where the phonon transport is most strongly suppressed with Eu 
doping). Figure 2 shows the field dependence of $\kappa_{ab}$ 
measured with magnetic field parallel to the $c$ axis or to the 
$ab$ plane for this La$_{1.98}$Eu$_{0.02}$CuO$_4$ single crystal. 
In both cases, there is no substantial magnetic field dependence 
up to 6 T at 12.5 K and 18.5 K. Therefore, magnetic origin for 
the strong phonon scattering is not likely. The last possible 
cause of the strong suppression of the phonon peak is the phonon 
scattering by the structural distortions introduced by Eu doping. 
It is known that a certain degree of Eu-doping in LCO is 
necessary to induce the structural phase transition from LTO to 
LTO2.\cite{Werner,Tsukada,note} When the doping level is too 
small to induce the macroscopic structural transition, it is most 
likely that the LTO2-like tilts of the CuO$_6$ octahedra exist 
locally around Eu$^{3+}$ ions in the background of LTO phase, 
which results in very strong lattice distortions that 
significantly scatter phonons. It should be noted that, since Sr 
doping can never induce phase transition from LTO to LTO2/LTT, 
the lattice distortions in La$_{1.98}$Sr$_{0.02}$CuO$_4$ are 
expected to be much weaker. 

\begin{figure} 
\includegraphics[clip,width=8.5cm]{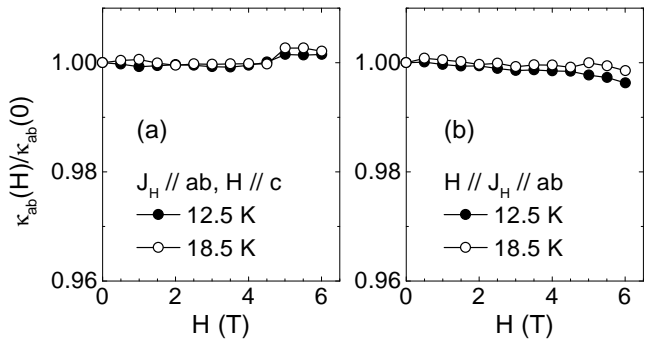} 
\caption{Magnetic field dependence of in-plane thermal 
conductivity of La$_{1.98}$Eu$_{0.02}$CuO$_4$ single crystal 
measured with magnetic field parallel to (a) the $c$ axis and (b) 
the $ab$ plane, respectively.} 
\end{figure} 

With increasing Eu concentration, the local LTO2 regions around 
Eu$^{3+}$ ions are expected to percolate at low temperature and 
the macroscopic structural transition takes place. The 
cooperative tilt of CuO$_6$ octahedra in LTO2 phase would weaken 
the lattice distortions and the phonon scattering, which is 
actually observed in La$_{1.8}$Eu$_{0.2}$CuO$_4$. As shown in 
Fig. 1(a), the phonon peak, although still very weak, reappears 
in La$_{1.8}$Eu$_{0.2}$CuO$_4$. 

The Eu-doping dependence of $\kappa_c$ shows that the effect of 
lattice distortion related to the LTO2 phase on the phonon 
transport is weaker in the $c$ axis than in the $ab$ plane. One 
can see in Fig. 1(b) that the phonon peak still exists in 
$\kappa_c$ of La$_{1.98}$Eu$_{0.02}$CuO$_4$, and the peak height 
is comparable to that in La$_{1.98}$Sr$_{0.02}$CuO$_4$ (where the 
disordering of static spin stripes in the $c$ axis causes rather 
strong phonon scattering\cite{Sun}). However, the temperature 
dependence of $\kappa_c$ in La$_{1.98}$Eu$_{0.02}$CuO$_4$ is too 
complicated (because of a double peak feature) to extract any 
detailed information of the phonon transport. Nevertheless, the 
$c$-axis phonon peak is larger in La$_{1.8}$Eu$_{0.2}$CuO$_4$ 
than in La$_{1.98}$Eu$_{0.02}$CuO$_4$, similar to that in the 
$ab$ plane. 

The above results show that in lightly Eu-doped LCO, the 
structural distortions associated with the local LTO2 regions 
strongly scatter $ab$-plane phonons. Such structural distortions 
and phonon scattering become weaker when the Eu concentration is 
large enough to stabilize the macroscopic LTO2 phase at low 
temperatures. However, compared to the pure LCO, the phonon peak 
is strongly suppressed even in La$_{1.8}$Eu$_{0.2}$CuO$_4$ which 
has global LTO2 phase. There are two possible reasons for such 
difference. First, it may come from the difference in phonon 
spectrum between LTO and LTO2 phases, which is, however, not very 
likely; as can be seen in Fig. 1, both $\kappa_{ab}$ and 
$\kappa_c$ of La$_{1.8}$Eu$_{0.2}$CuO$_4$ show a step-like 
increase at $\sim$ 130 K when the structure changes from LTO to 
LTO2 phase, which means the LTO2 phase essentially has better 
phonon heat transport than the LTO phase. Another, more likely 
reason is simply the impurity scattering by Eu dopants. 

\subsection{Magnetic properties of La$_{2-y}$Eu$_y$CuO$_4$}

It is well known that the magnetic properties of La$_2$CuO$_4$ 
are determined by the Cu spins. At high temperatures, LCO is 
essentially a two-dimensional Heisenberg antiferromagnet and a 
weak interlayer coupling gives rise to the three-dimensional 
long-range N\'{e}el order.\cite{Kastner,Keimer} In the LTO-phase 
of LCO, the spin easy axis is the $b$ axis and all the spins are 
weakly canted along the $c$ axis. Such canted moments depend on 
the tilting of CuO$_6$ octahedra; they become weaker in the LTO2 
phase and disappear in the LTT phase.\cite{Tsukada} Thus, it is 
helpful to study the magnetic properties for understanding the 
Eu-doping effect on local structure. Figure 3 shows the magnetic 
susceptibility data of Eu-doped LCO single crystals measured with 
5000 Oe field applied along the $c$ axis or the $ab$ plane. The 
pure LCO sample shows a sharp N\'{e}el transition in $\chi_c$ and 
a much broader peak in $\chi_{ab}$ at $T_N$ = 307 K, which 
originates from the AF ordering of Cu$^{2+}$ canted 
moments.\cite{Thio1,Thio2,Lavrov2} 

\begin{figure} 
\includegraphics[clip,width=8.5cm]{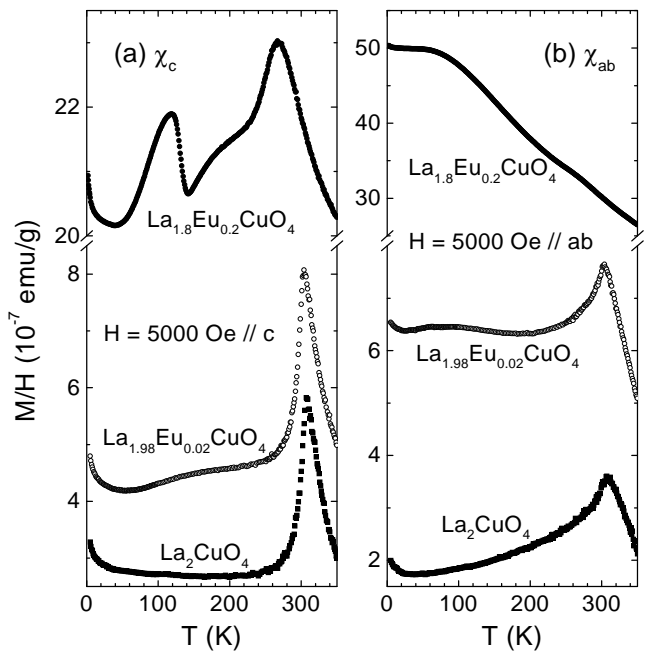} 
\caption{Magnetic susceptibility of La$_{2-y}$Eu$_y$CuO$_4$ 
single crystals measured in 5000 Oe field applied along (a) the 
$c$ axis and (b) the $ab$ plane.} 
\end{figure} 

Eu-doping induces pronounced changes in the magnetic properties 
of La$_{1.8}$Eu$_{0.2}$CuO$_4$, as shown in Fig. 3. First, both 
$\chi_c$ and $\chi_{ab}$ data are shifted up significantly. This 
additional signals apparently come from the Van Vleck 
contribution of Eu$^{3+}$ ions, which is weakly $T$-dependent in 
$\chi_c$ and noticeably $T$-dependent in $\chi_{ab}$, 
respectively. (Note that the Van Vleck term, although is 
independent of temperature in usual cases, can be $T$-dependent 
when the energy difference between the ground state and the 
excited orbital state is smaller than $k_{B}T$, as shown for 
Eu$_2$CuO$_4$.\cite{Tovar}) Second, a clear N\'{e}el transition 
shows up in $\chi_c$ with lower $T_N$ (268 K) compared to undoped 
LCO, while the corresponding peak is almost smeared out in 
$\chi_{ab}$ (only a slight hump at $T_N$) because of the strong 
$T$-dependence of $\chi_{ab}$. Third, another transition appears 
in $\chi_c$ at the structural transition temperature, which is 
due to the suppression of canted moments in the LTO2 phase and 
the reduction of the interlayer magnetic coupling.\cite{Tsukada} 
It is not clear whether there is a corresponding transition in 
$\chi_{ab}$ because of its strong $T$-dependence. The details of 
the magnetic structure in La$_{1.8}$Eu$_{0.2}$CuO$_4$ are 
thoroughly discussed elsewhere.\cite{Tsukada} It should be 
pointed out that the susceptibility data of the present single 
crystal are different from the previous report, which used 
polycrystalline samples and claimed that 
La$_{1.8}$Eu$_{0.2}$CuO$_4$ has the LTT structure.\cite{Kataev}  

For Eu-1\%-doping, the susceptibility data show moderate changes. 
First, the enhancement of both $\chi_c$ and $\chi_{ab}$ are much 
smaller than those in La$_{1.8}$Eu$_{0.2}$CuO$_4$. Second, 
N\'{e}el transition is only slightly shifted to a lower 
temperature ($T_N$ = 304 K). Thus, the high temperature magnetic 
properties do not show any drastic change upon Eu-doping, which 
is understandable because the structural changes associated with 
Eu doping only happen at low temperatures. This is also 
consistent with the weak change of the high-$T$ magnon peak in 
$\kappa_{ab}$. (In contrast, the magnon peak disappears in 
La$_{1.98}$Sr$_{0.02}$CuO$_4$ (Fig. 1), in which the N\'{e}el 
transition also disappears.\cite{Keimer, Sun}) Third, there is no 
second transition in $\chi_c$ at $\sim$ 130 K, because there is 
no structural phase transition in this lightly-doped sample. 
However, $\chi_c$ decreases slowly with deceasing temperature 
from $\sim$ 150 K to 50 K and $\chi_{ab}$ shows a weak hump at 
the same temperature region. This is in correspondence with our 
picture that slight Eu-doping induces local LTO2 regions in the 
LTO background, although no macroscopic structural transition 
occurs.  

\subsection{Phonon peak in La$_{1.68}$Eu$_{0.2}$Sr$_{0.12}$CuO$_4$} 

\begin{figure} 
\includegraphics[clip,width=6.5cm]{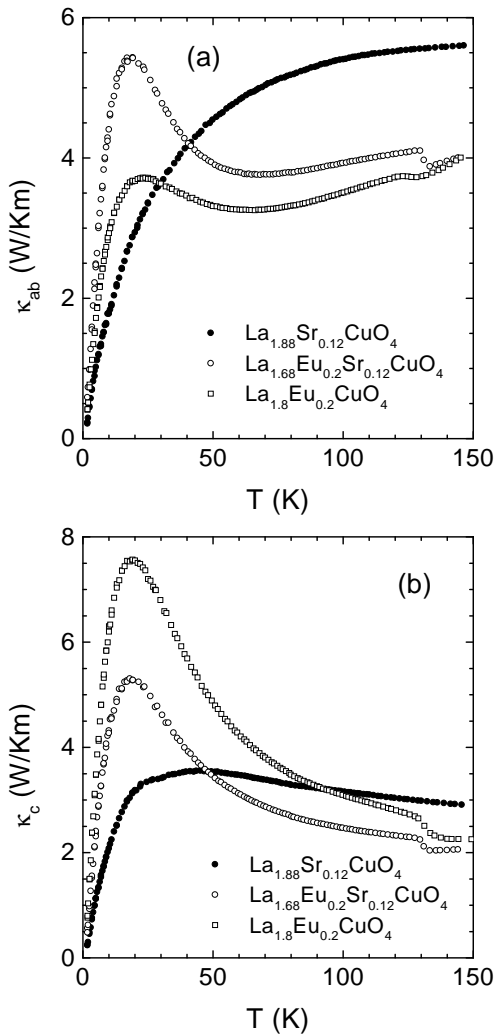} 
\caption{Low-temperature thermal conductivity of 
La$_{1.88}$Sr$_{0.12}$CuO$_4$ and 
La$_{1.68}$Eu$_{0.2}$Sr$_{0.12}$CuO$_4$ single crystals. The data for 
La$_{1.8}$Eu$_{0.2}$CuO$_4$ are also shown for comparison. The 
step-like feature at $\sim$ 130 K is due to the LTO $\to$ LTT and LTO 
$\to$ LTO2 structural transition for 
La$_{1.68}$Eu$_{0.2}$Sr$_{0.12}$CuO$_4$ and 
La$_{1.8}$Eu$_{0.2}$CuO$_4$, respectively.} 
\end{figure} 

While Eu-doping significantly suppresses the phonon peak in the 
non-carrier-doped LCO, it shows quite different effects on the 
phonon heat transport of LSCO. The $\kappa_{ab}(T)$ and 
$\kappa_c(T)$ data for La$_{1.88}$Sr$_{0.12}$CuO$_4$ and 
La$_{1.68}$Eu$_{0.2}$Sr$_{0.12}$CuO$_4$ single crystals are shown 
in Fig. 4, in which the data for La$_{1.8}$Eu$_{0.2}$CuO$_4$ are 
also included for comparison. Since there are certain amount of 
charge carriers doped by Sr, we should separate the electronic 
contribution from the total thermal conductivity before 
discussing the mechanism for the phonon heat transport. 

The electronic thermal conductivity $\kappa_e$ and the electrical 
resistivity $\rho$ are related by $\kappa_e=LT/\rho$, where $L$ 
is called the Lorenz number. In simple metals, $L$ is usually 
constant at high-$T$ and low-$T$ (the Wiedemann-Franz law) and is 
given by the Sommerfeld value $2.44 \times 10^{-8}$ 
W$\Omega$/K$^2$. When the electron-electron correlation becomes 
strong, $L$ becomes smaller. For YBa$_2$Cu$_3$O$_{7-\delta}$, $L$ 
has been estimated to be $1.2-2.0 \times 10^{-8}$ W$\Omega$/K$^2$ 
near $T_c$ by Hirschfeld and Putikka,\cite{Hirschfeld} while 
Takenaka {\it et al.} \cite{Takenaka} estimated $L$ to be 
$2.4-3.3 \times 10^{-8}$ W$\Omega$/K$^2$ above $T_c$. So there is 
still no consensus value of $L$ for cuprates. Here, we roughly 
estimate $\kappa_e$ of our crystals by using the Sommerfeld 
value. The in-plane resistivity $\rho_{ab}$ of 
La$_{1.88}$Sr$_{0.12}$CuO$_4$ and 
La$_{1.68}$Eu$_{0.2}$Sr$_{0.12}$CuO$_4$ are measured using a 
standard ac four-probe method and shown in the inset of Fig. 5. 
We note that the zero resistance at 5 K in 
La$_{1.68}$Eu$_{0.2}$Sr$_{0.12}$CuO$_4$ is a filamentary 
superconducting effect, since the dc susceptibility measurement 
does not show any bulk superconductivity in this sample. The main 
panel of Fig. 5 shows the estimated in-plane phononic thermal 
conductivity $\kappa_{ph,ab}$ ($ =\kappa_{ab}-\kappa_{e,ab}$). It 
can be seen that the main contribution to thermal conductivity is 
phononic in these samples. For the $c$-axis heat transport, the 
electronic contribution is negligibly small because of the 2 -- 3 
orders of magnitude larger electrical resistivity in the $c$ axis 
compared to $\rho_{ab}$.\cite{Nakamura2,Nakamura3} The 
experimental data for $\kappa_c$ is nearly pure phonon 
conductivity. 

\begin{figure} 
\includegraphics[clip,width=6.5cm]{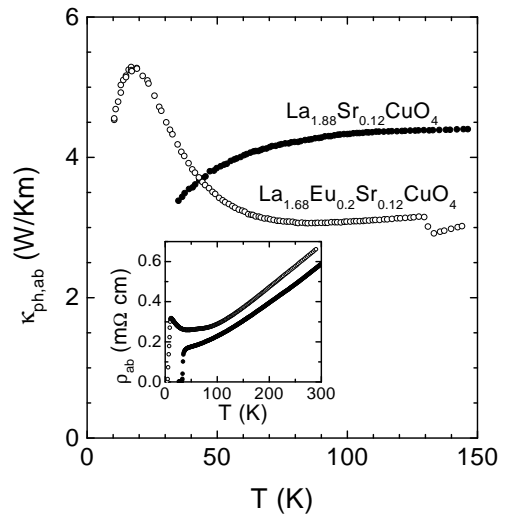} 
\caption{Comparison of the in-plane phononic thermal conductivity 
$\kappa_{ph,ab}$ of La$_{1.88}$Sr$_{0.12}$CuO$_4$ single crystal 
to that of La$_{1.68}$Eu$_{0.2}$Sr$_{0.12}$CuO$_4$ crystal; the 
electronic thermal conductivity $\kappa_{e,ab}$ is estimated from 
the in-plane resistivity data by the Wiedemann-Franz law and 
$\kappa_{ph,ab} = \kappa_{ab}-\kappa_{e,ab}$. Inset: in-plane 
resistivity data for La$_{1.88}$Sr$_{0.12}$CuO$_4$ (solid 
circles) and La$_{1.68}$Eu$_{0.2}$Sr$_{0.12}$CuO$_4$ (open 
circles).} 
\end{figure} 

Clearly, the phonon heat transport is strongly damped in 
La$_{1.88}$Sr$_{0.12}$CuO$_4$, which shows complete disappearance 
of the phonon peak in $\kappa_{ab}$ and a weak hump in 
$\kappa_c$. The strong phonon scattering can be related to the Sr 
dopants acted as impurities, the charge carriers, and the 
structural distortions associated with dynamical stripes. 
Eu-doping in this LSCO system significantly increases the low-$T$ 
phononic thermal conductivity; clear phonon peak reappears in 
both $\kappa_{ab}$ and $\kappa_c$. Similar behavior in R-doped 
LSCO was first attributed by Baberski {\it et al.}\cite{Baberski} 
to the weakening of phonon scattering in the static stripe phase. 
However, their data were collected on polycrystalline samples and 
did not display quantitative difference between LSCO and R-doped 
LSCO. Here the data on single crystals show convincing evidence 
for the importance of static stripes on the phonon transport. In 
this Eu-doped LSCO, LTO $\to$ LTT structural transition takes 
place at $\sim$ 130 K [Ref. \onlinecite{Klauss,Werner}] and 
results in a step-like increase of the phononic thermal 
conductivity in the LTT phase. An important feature is, as shown 
in Figs. 4(b) and 5, that La$_{1.88}$Sr$_{0.12}$CuO$_4$ has 
larger phononic thermal conductivity than 
La$_{1.68}$Eu$_{0.2}$Sr$_{0.12}$CuO$_4$ in both the $ab$ plane 
and the $c$ axis for $T >$ 45 K, which means 
La$_{1.68}$Eu$_{0.2}$Sr$_{0.12}$CuO$_4$ has stronger phonon 
scattering (which may come from Eu dopants) at high temperatures 
even in the LTT phase. Therefore, it is more reasonable to 
attribute the reappearance of phonon peak in Eu-doped LSCO to the 
stabilization of stripes at low temperatures, which reduces the 
strong phonon scattering by the stripe fluctuations, rather than 
the difference of the phonon heat transport between LTT and LTO 
phases. 

By re-examining the previous thermal conductivity data for 
La$_{1.28}$Nd$_{0.6}$Sr$_{0.12}$CuO$_4$ single crystal,\cite{Sun} 
we find that the above feature is also present in Nd-doped LSCO. 
This means that the suppression of the phonon scattering is 
common to the R-doped LSCO, where static charge stripes are 
formed. 

Furthermore, by comparing the data of 
La$_{1.68}$Eu$_{0.2}$Sr$_{0.12}$CuO$_4$ with those of 
La$_{1.8}$Eu$_{0.2}$CuO$_4$, which has LTO2 phase without 
stripes, we can obtain useful information on the scattering 
mechanism of phonons. In Figs. 4(a) and 5, one can see that in 
the $ab$ direction the phonon peak in 
La$_{1.68}$Eu$_{0.2}$Sr$_{0.12}$CuO$_4$ is larger than that in 
La$_{1.8}$Eu$_{0.2}$CuO$_4$, although the former compound has 
much more dopants and charge carriers. This difference clearly 
shows that the LTT phase has stronger phonon transport than the 
LTO2 phase, which overcomes the additional impurity-phonon 
scattering by Sr dopants and charge-phonon scattering in 
La$_{1.68}$Eu$_{0.2}$Sr$_{0.12}$CuO$_4$. However, the phonon peak 
in $\kappa_c$ shows opposite trend, that is, the peak magnitude 
is larger in La$_{1.8}$Eu$_{0.2}$CuO$_4$ than in 
La$_{1.68}$Eu$_{0.2}$Sr$_{0.12}$CuO$_4$. The additional phonon 
scattering in the $c$ axis of 
La$_{1.68}$Eu$_{0.2}$Sr$_{0.12}$CuO$_4$ is unlikely due to the Sr 
dopants, considering their negligible effect on $\kappa_{ab}$. 
Instead, it is probably due to the lattice disorder induced by 
the disordering of static stripes in the $c$ direction.\cite{Sun} 
There are two factors that make the stripe correlation in the $c$ 
direction very weak even in the static stripe phase. First, the 
neutron experiments have already shown\cite{Tranquada2} that the 
magnetic correlation length in the $c$ direction is very short. 
Second, the stripe orientations have been proposed to rotate 
90$^{\circ}$ from one CuO$_2$ plane to the nearest-neighbor 
plane.\cite{Tranquada2} Such strong disorder of the static 
stripes along the $c$ axis leads to rather strong phonon 
scattering in the $c$-axis transport, which is very similar to 
what was observed in the lightly-doped LSCO.\cite{Sun}  

\section{Summary} 

We have measured the $ab$-plane and the $c$-axis thermal 
conductivities of La$_{2-y}$Eu$_y$CuO$_4$ ($y$ = 0, 0.02 and 0.2) 
and La$_{1.88-y}$Eu$_y$Sr$_{0.12}$CuO$_4$ ($y$ = 0 and 0.2) 
single crystals. It is found that the low-temperature phonon heat 
transport shows opposite Eu-doping dependence in these two 
systems, that is, Eu-doping strongly suppresses the phonon peak 
in LCO, while it induces the reappearance of phonon peak in LSCO. 
In Eu-1\%-doped LCO, the phonon peak in $\kappa_{ab}$ is 
anomalously wiped out, and such strong phonon scattering is 
caused by the lattice distortions induced by the local LTO2-like 
regions around Eu ions in the LTO background. Increasing 
Eu-doping in LCO to 10\% leads to the LTO $\to$ LTO2 structural 
transition, which reduces the lattice distortion and phonon 
scattering. The phonon peak, although observed in 
La$_{1.8}$Eu$_{0.2}$CuO$_4$, is still much smaller than that in 
pure LCO. On the other hand, Eu-doping in LSCO enhances the 
phonon heat transport, which is likely due to the formation of 
static stripes that reduces the phonon scattering. Comparison of 
the phonon heat transport between 
La$_{1.68}$Eu$_{0.2}$Sr$_{0.12}$CuO$_4$ and 
La$_{1.8}$Eu$_{0.2}$CuO$_4$ tells us that phonon scattering in 
the $c$ axis is rather strong even in the static stripe phase, 
which is consistent with the fact that the stripes are not well 
ordered in the $c$ axis. 

\begin{acknowledgments} 
We thank J. Takeya for technical assistance, A. N. Lavrov and I. 
Tsukada for helpful discussions.  
\end{acknowledgments}


\begin{thebibliography}{}
\bibitem{note} 
It was reported in Ref. \onlinecite{Werner} that 
La$_{1.8}$Eu$_{0.2}$CuO$_4$ has LTT phase, while the new result 
on single crystal (Ref. \onlinecite{Tsukada}) suggests that this 
compound has LTO2 phase, similar to Nd-doped LCO (Ref. 
\onlinecite{Crawford}).
\end{thebibliography}
\end{document}